\documentclass{ws-mpla}

\begin{document}

\markboth{Dong-han Yeom, Sungwook E. Hong and Heeseung Zoe}
{Critical Reviews of causal patch Measure over the Multiverse}

%%%%%%%%%%%%%%%%%%%%% Publisher's Area please ignore %%%%%%%%%%%%%%
\catchline{}{}{}{}{}
%%%%%%%%%%%%%%%%%%%%%%%%%%%%%%%%%%%%%%%%%%%%%%%%%%%%%%%%%%%%%%%%%%%

\title{CRITICAL REVIEWS OF CAUSAL PATCH MEASURE \\
OVER THE MULTIVERSE
\footnote{A proceeding for CosPA2008. Talk on the 29th of October, 2008, Pohang, Korea.}}

\author{\footnotesize DONG-HAN YEOM${}^\dag$, SUNGWOOK E.~HONG${}^\ddag$ and HEESEUNG ZOE${}^\S$}

\address{Department of Physics, KAIST,\\
Daejeon 305-701, South Korea \\
${}^\dag$innocent@muon.kaist.ac.kr\\
${}^\ddag$eostm@muon.kaist.ac.kr\\
${}^\S$heeseungzoe@iyte.edu.tr}

\maketitle

%\pub{Received (Day Month Year)}{Revised (Day Month Year)}

\begin{abstract}
In this talk, the causal patch measure based on black hole complementarity is critically reviewed.
By noticing the similarities between the causal structure of an inflationary dS space and that of a black hole,
we have considered the complementarity principle between the inside and the outside of the causal horizon as an attractive way 
to count the inflationary multiverse.
Even though the causal patch measure relieves the Boltzmann brain problem and stresses physical reality based on observations,
it could be challenged by the construction of counterexamples, both on regular black holes and charged black holes, to black hole complementarity.

\end{abstract}

\ccode{PACS Nos.: include PACS Nos.}

\section{Introduction}

If the paradigm of inflation is correct, then there would be no other option than to conclude that
the global structure is the multiverse, a complex mixture of thermalized and inflating regions.
In the context of the multiverse, we could explore the cosmic fine-tuning problem by asking for statistical distributions over the multiverse.
An approach based on the statistical distribution is deeply related to anthropic reasoning.
Whether our universe is typical among the possible pocket universes would be an important signature
of the statistical inference and the reasoning of quantum cosmology.

In general, the complexity of the global structure of the multiverse depends on the shape of inflation potentials.
However, Guth pointed out that eternal inflation could be activated even with a simple potential,
and there would be no proper way to quantify the statistical distribution 
based on the volume fractions due to the infinity and the slice-dependence \cite{guth}.
Thus it implies that we may fail to handle the complexity and to specify the measure over the inflationary multiverse.
This problem is called the `measure problem.'

Among many proposals to overcome the measure problem, there have been a remarkable category of attempts inspired by
the similarities between physical properties (e.g. causal structures, thermodynamics, etc.) of black holes and those of de Sitter (dS) spaces.
Susskind proposed that the complementarity principle should be involved in specifying the measure
from his speculations for resolving the black hole information paradox \cite{susskind} \cite{complementarity}.
In the line of this thought, Bousso has developed a causal patch measure based on black hole complementarity \cite{bousso}.

\section{Black Hole Complementarity and causal patch Measure}

\subsection{Black Hole Complementarity}

According to black hole complementarity,
there is no consistent way of describing both the inside and the outside of the black hole horizon \cite{complementarity}.
If we take a free-falling observer who steps into the horizon, 
he would investigate the internal structure of the black hole and confirm the original information
which we used to make the black hole.
If we take an additional asymptotic observer who stays outside of the horizon,
he can reconstruct the original information from Hawking radiations by considering correlations among them.
From a heuristic argument using the statistical average by tracing out the black hole internal states,
Page suggested that Hawking radiation begins to carry the information only after so called ``information retention time,''
which is the moment when the size of the black hole horizon decreases into half \cite{page}.

\begin{figure}[hbt]
\begin{center}
\psfig{file=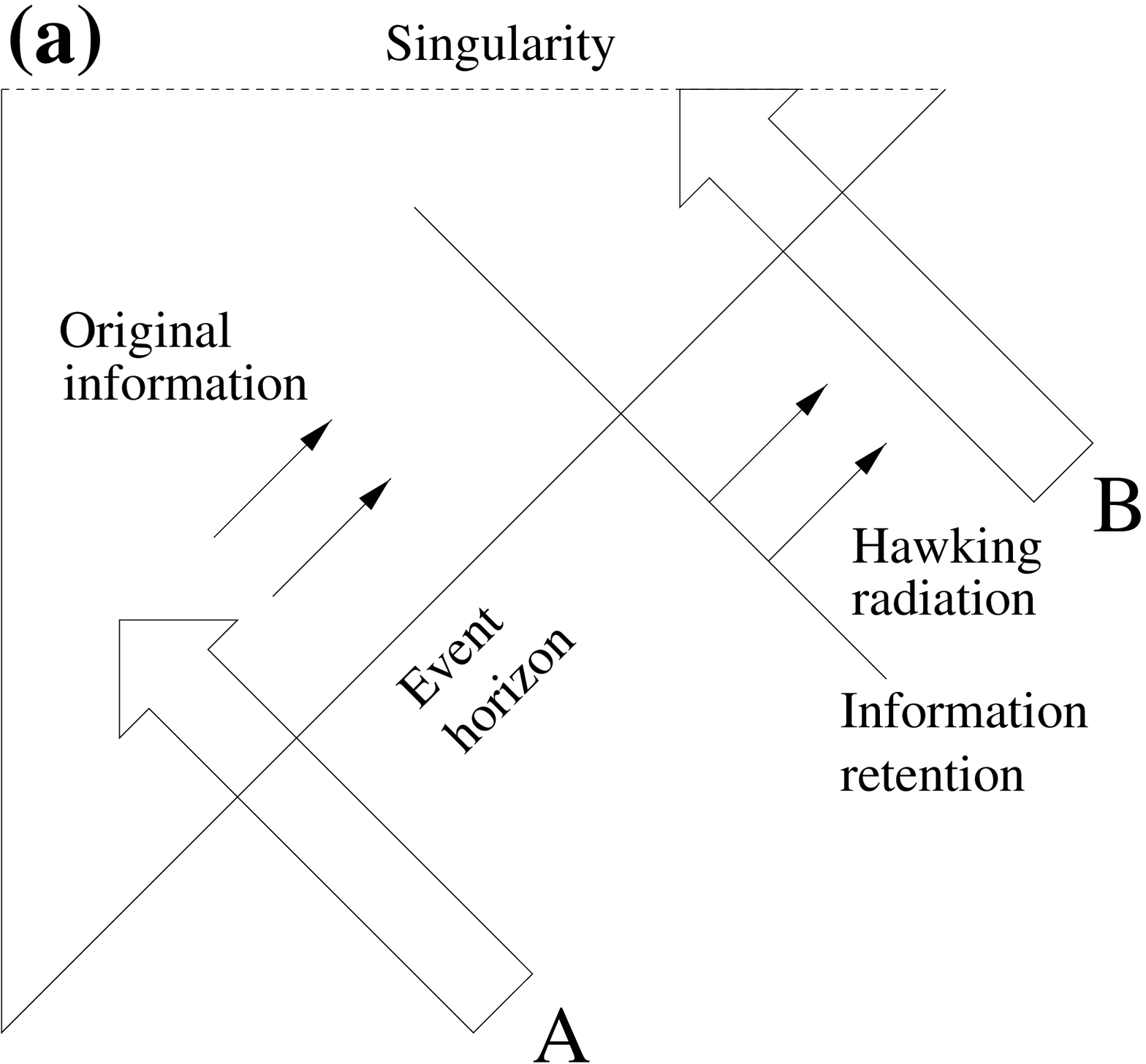,scale=.25}
\hspace*{50pt}
\psfig{file=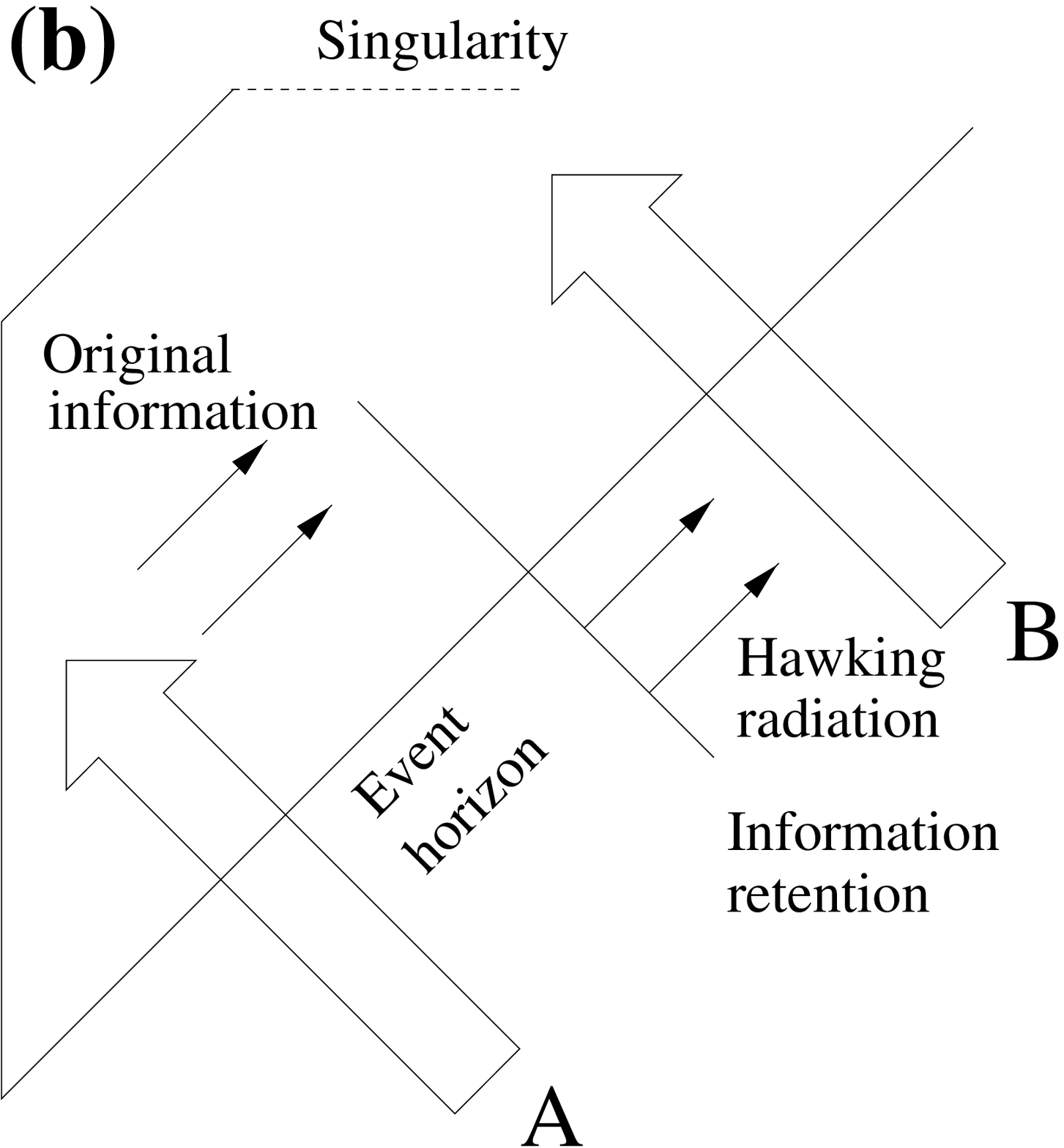,scale=.25}
\end{center}
\vspace*{8pt}
\caption{Possibilities of duplication experiments in different types of black holes: 
(a) duplication experiment is not possible in a Schwartzschild black hole,
due to the lack of time in the presence of a singularity inside the black hole.
(b) It may be possible to find or construct a black hole geometry in which we have enough time to do the duplication experiment.}
\label{fig1}
\end{figure}

%There seem to be two pieces of the same information.
Essentially, black hole complementarity assumes that all information is duplicated on the stretched horizon.
So now we have actually the duplication of quantum information, which is an obvious violation of quantum principles.
However, these two observers cannot communicate by quanta whose energy is less than the total black hole mass energy,
so this complementary viewpoint is logically safe and operationally meaningful.
This is quite true at least in Schwarzschild black holes.
Let us consider two observers A and B in \figurename ~\ref{fig1}(a).
At first, send an observer A to investigate the inner states of the Schwarzschild black hole and identify the original information 
about the resources that made the black hole.
After the information retention time of the order of the black hole mass $M$,
send an observer B to check whether the information is really duplicated.
We can calculate the time for observer B to hit the singularity at $X_{-}X_{+}=M^2$ in the Kruskal coordinate,
which is $M^2\exp \left(-M^2\right)$: 
If observer A sends his information to observer B, he could do it only by sending quanta
whose energy is on the order of $M^{-2}\exp \left(M^2\right)$ which is much greater than the total mass energy.
This means that observer B can never confirm that the quantum information is duplicated.

\subsection{Causal Patch Measure}

If the complementarity principle is a fundamental principle of quantum gravity, it should be considered
even when we try to specify the measure over the inflationary multiverse, 
because the inflationary de Sitter universe has its own causal horizon.
According to complementarity, there must be no way to involve the causal parts outside of the observer horizon in the physical description.
This point makes the reasoning of causal patch measure economical and plausible, 
as it focuses on only the causally connected parts of the multiverse and abandons the outside of the cosmic horizon.

There are two major merits in the causal patch measure.
First, the causal patch measure makes the issues related with typicality simple:
Since complementarity restricts the discussion only within the cosmic horizon,
the causal patch measure can relieve the Youngness paradox and the Boltzmann brain problem,
which are evoked by considering the whole global structure of the multiverse.
Second, the causal patch measure makes the multiverse testable by observations\textit{ in principle}:
Although the whole multiverse is still always beyond our ability of observation,
the concept of the multiverse can be tested and modified by observations,
since complementarity, as a fundamental principle, protects the logic that neglects magnificent parts of the global structure.

\section{Looking for Counterexamples to Complementarity}

In \figurename ~\ref{fig1}(a), the main reason that black hole complementarity is working in a Schwarzschild black hole is
the lack of time to check whether the information is duplicated and so the no-cloning theorem is violated.
Therefore we can ask whether we can find or generate black hole geometries to have enough time to do
the duplication experiment.
The authors claim that we can find such counterexamples to black hole complementarity and that they can also challenge
the causal patch measure based on these counterexamples.

\begin{figure}[hbt]
\begin{center}
\psfig{file=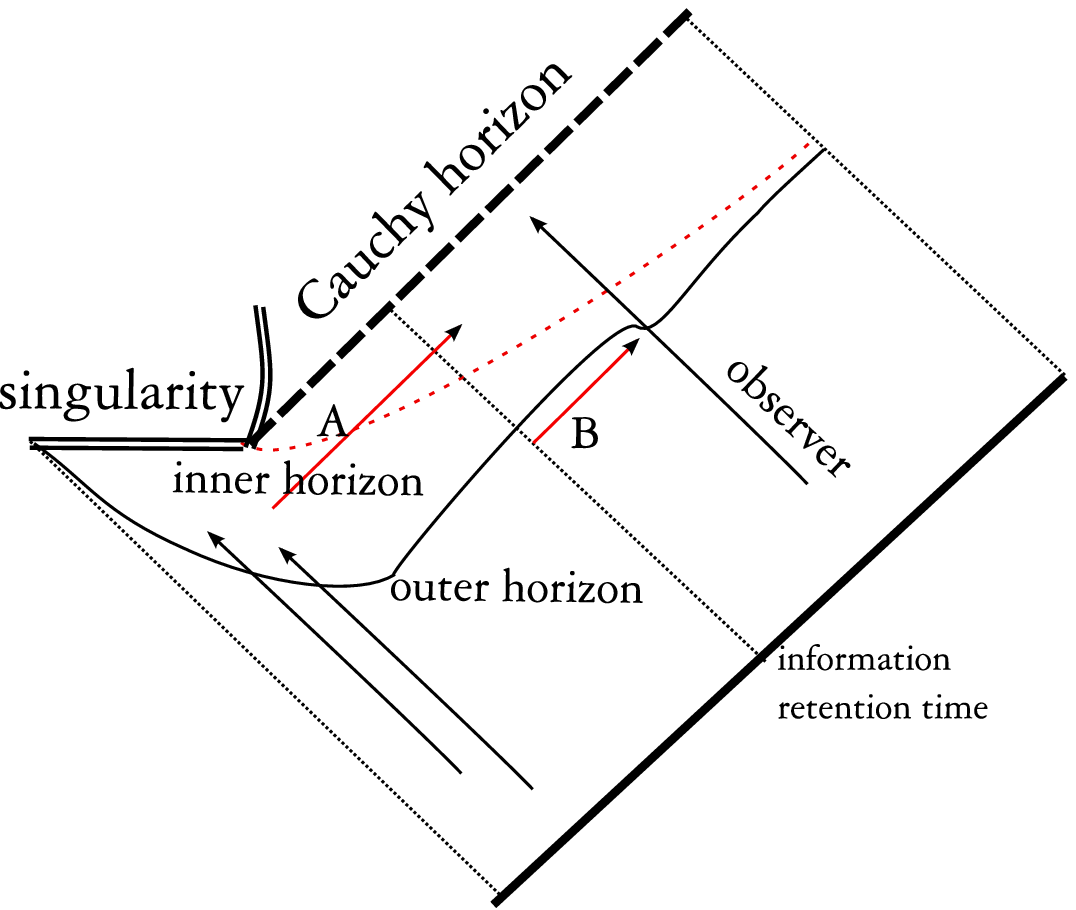,scale=.45}
\hspace*{50pt}
\psfig{file=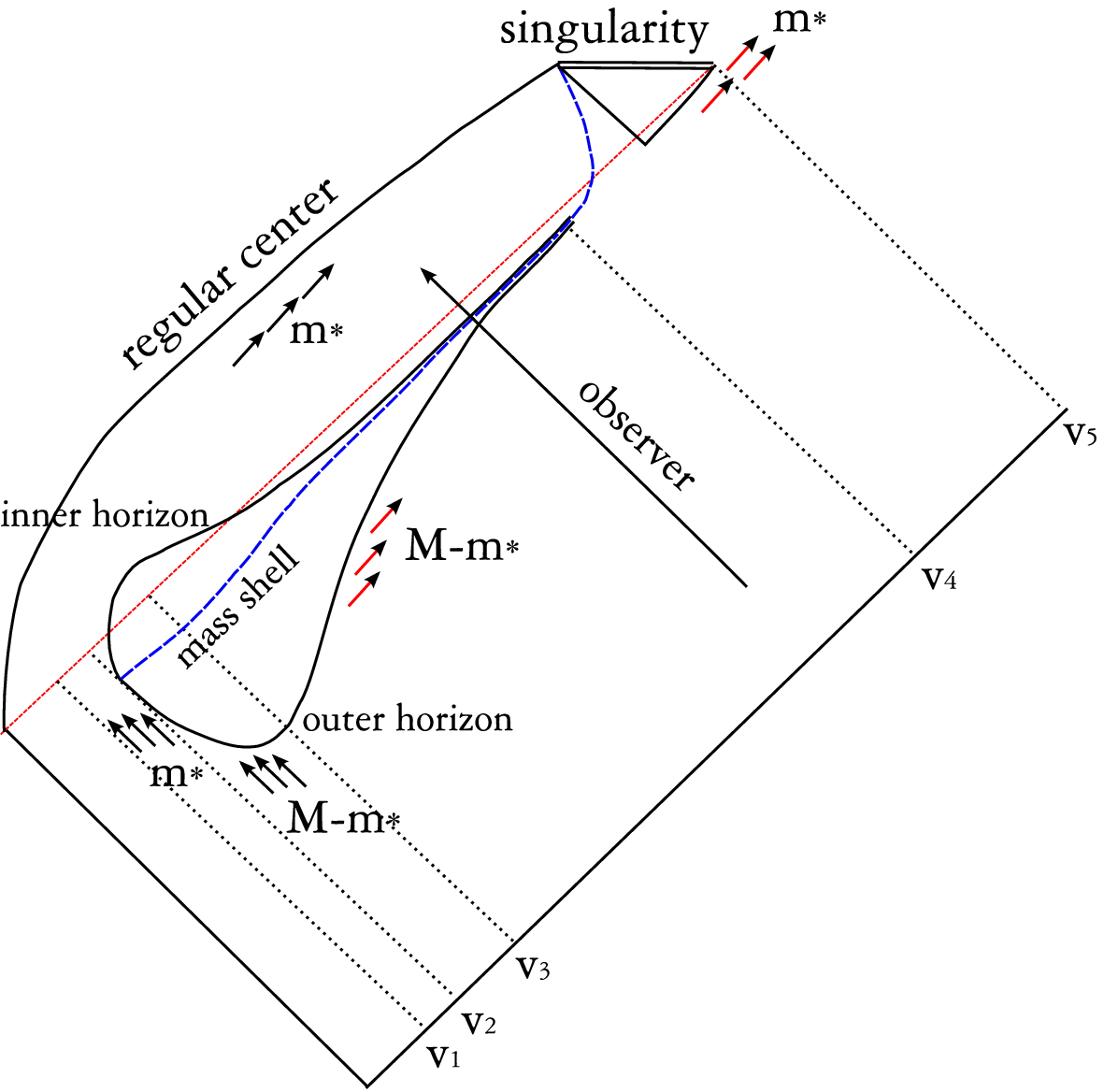,scale=.45}
\end{center}
\vspace*{8pt}
\caption{Causal structures of a dynamical charged black hole (left)
and Frolov, Markov, and Mukhanov's regular black hole (right).}
\label{fig2}
\end{figure}

\subsection{Charged Black Holes}

The evolution of a charged black hole is dynamically described in Ref.~\refcite{HHSY}.
By assuming a complex massless scalar field coupled with an electromagnetic field,
we can effectively include Hawking radiation and get the Penrose diagram as the result of numerical simulations 
(see the left panel of \figurename ~\ref{fig2}).
There exists a singular region over the Cauchy horizon,
and the region near the inner horizon due to mass inflation also seems to be singular.
However, by increasing the number of massless scalar fields,
we can obtain a regular region between the inner horizon and the Cauchy horizon.
This means that we can have enough time to do the duplication experiment,
and so black hole complementarity is not correct in this charged black hole any longer.

Still there is a loophole in the argument:
If there is a non-local effect in Hawking radiation, as Horowitz and Maldacena proposed,
then the duplication experiment may not be possible,
because Hawking radiation can carry the information only after the matter states are projected
at the singularity near the Cauchy horizon \cite{HM}.
However, if we consider randomized mixing of quantum states on the stretched horizon,
we can get rid of this loophole.
After Hawking radiation comes out from a charged black hole,
if we add one bit of quantum information, then it will come out as Hawking radiations
in the time of order $M\ln M$.
Thus right after this charged black hole acts like an information mirror,
the duplication experiment is possible \cite{HP}.

\subsection{Regular Black Holes}

It is possible to construct a counterexample with regular black holes
whose inside structure is a de Sitter space but whose outside is a Schwarzschild black hole 
(see the right panel of \figurename ~\ref{fig2}) \cite{regular}.
The cosmological constant of the de Sitter space is coupled with a scalar field, 
which decides the dynamical evolution of the regular black hole.
The scalar field tunnels into the greater vacuum energy
and then the regular black hole goes to the extreme limit.
(This means that the tunneling events produce statistical ensembles
and that some of them may satisfy the criteria for the duplication experiment).
In this case, we have enough time to do the duplication experiment
between the information retention time and, surely, the extreme limit.

\subsection{Inflationary de Sitter Spaces}

By the previous counterexamples, black hole complementarity seems to be invalid in general,
which shows that the causal patch measure has unstable ground for its consistency.
According to the authors' knowledge, complementarity in de Sitter spaces is more difficult to realize than it is in black holes,
since even physical concepts like particles are not well-defined in de Sitter spaces.

However, it is also difficult to invalidate dS complementarity directly by constructing counterexamples
in which we do the duplication experiment:
The information retention time of the de Sitter space is estimated to be as much as the Poincare recurrence time \cite{dsinfo}.
This means that if we want to observe dS (Hawking) radiation with information,
we should wait for the whole lifetime of dS space.
This may be good news for advocates of the causal patch measure,
for they would not be troubled with counterexamples using duplication experiments.

\section{Conclusion}

Black hole complementarity is a fascinating idea intended to resolve the black hole information puzzle.
The causal patch measure based on black hole complementarity has some nice features to the measure problem of the inflationary multiverse:
to relieve the issues around the typicality and to relate the discussions directly to observational knowledge.
However, black hole complementarity seems to be invalidated by counterexamples permitting duplication experiments
in charged black holes and regular black holes,
which critically affect the consistency of the causal patch measure.

\end{document}